\documentclass{article}
\usepackage[letterpaper, portrait, margin=1in]{geometry}
\usepackage{graphicx}
\usepackage{amsmath, bm}
\usepackage{amsfonts}
\usepackage{amssymb}
\usepackage{amsthm} % theorem package
\usepackage[utf8]{inputenc}
\usepackage{amsthm}
\usepackage[english]{babel}
\usepackage{hyperref}

\makeatletter
\renewcommand{\@seccntformat}[1]{}
\makeatother

\usepackage{sectsty}
\sectionfont{\centering}

\begin{document}
\title{Convolutional Neural Network Achieves Human-level Accuracy in Music Genre Classification}
\author{Mingwen Dong\\Psychology, Rutgers University (New Brunswick)\\ mingwen.dong@rutgers.edu}
\date{}
\maketitle

\begin{abstract}
Music genre classification is one example of content-based analysis of music signals. Traditionally, human engineered features were used to automatize this task and 61\% accuracy has been achieved in the 10-genre classification. However, it's still below the 70\% accuracy that humans could achieve  in the same task. Here, we propose a new method that combines knowledge of human perception study in music genre classification and the neurophysiology of the auditory system. The method works by training a simple convolutional neural network (CNN) to classify a short segment of the music signal. Then, the genre of a music is determined by splitting it into short segments and then combining CNN's predictions from all short segments. After training, this method achieves human-level (70\%) accuracy and the filters learned in the CNN resemble the spectrotemporal receptive field (STRF) in the auditory system \footnote{All codes are available at: \url{https://github.com/ds7711/music_genre_classification}}.

\end{abstract}

\section{Introduction}

With the rapid development of digital technology, the amount of digital music content increases dramatically everyday. To give better music recommendations for the users, it's essential to have an algorithm that could automatically characterize the music. This process is called Musical Information Retrieval (MIR) and one specific example is music genre classification. 

However, music genre classification is a very difficult problem because the boundaries between different genres could be fuzzy in nature. For example, testing with a 10-way forced choices task, college students could achieve 70\% classification accuracy after hearing 3 seconds of the music and the accuracy doesn't improve with longer music \cite{tzanetakis2002musical}. Also, the number of labeled data often is much smaller than the dimension of the data. For example, \href{http://marsyasweb.appspot.com/download/data_sets/}{GTZAN dataset} \footnote{Available at: \url{http://marsyasweb.appspot.com/download/data_sets/}} used in the current work contains only 1000 audio tracks, but each audio track is 30s long with a sampling rate 22,050 Hz.  

Traditionally, using human-engineered features like MFCC (Mel-frequency cepstral coefficients), texture, beat and so on, 61\% accuracy has been achieved in the 10-genre classification task \cite{tzanetakis2002musical}. More recently, using PCA-whitened spectrogram as input, convolutional deep belief network has achieved 70\% accuracy in a 5-genre classification task. These results are reasonable but still not as good as humans, suggesting there's still space to improve.

Psychophysics and physiology study show that human auditory system works in a hierarchical way \cite{schnupp2011auditory}.  First, the ear decomposes the continuous sound waveform into different frequencies with higher precision on low frequencies. Then, neurons from lower to higher auditory structures gradually extract more complex spectro-temporal features with more complex spectro-temporal receptive field (STRF) \cite{theunissen2014neural}. The features used by human auditory system for music genre classification probably rely on these STRFs. By having the spectrogram as input and the corresponding genre as label, CNN will learn filters that extract features in the frequency and time domain. If these learned filters mimic STRFs in the human auditory system, they can extract useful features for music genre classification. Because music signal often is high-dimension in the time domain, having a CNN that fits the complete spectrogram of the music signal is not feasible. To solve this problem, we used a "divide-and-conquer" method: split the spectrogram of the music signal into consecutive 3-second segments, make predictions for each segment, and finally combine the predictions together. The main rational for this method is that humans' classification accuracy plateaus at 3 seconds and good results were obtained using 3-second segments to train convolutional deep belief network \cite{tzanetakis2002musical} \cite{lee2009unsupervised}. It also intuitively makes sense because different parts of the same music probably should belong to the same genre. 

To further reduce the dimension on the spectrogram, we used mel-spectrogram as the input to the CNN. Mel-spectrogram approximates how human auditory system works and can be seen as the spectrogram smoothed in the frequency domain, with high precision in the low frequencies and low precision in the high frequencies \cite{o1987speech} \cite{picone1993signal}.

\section{Data Processing \& Models}

\subsection{Data pre-processing}
\begin{figure}
	\centering
	\includegraphics[width=0.7\linewidth]{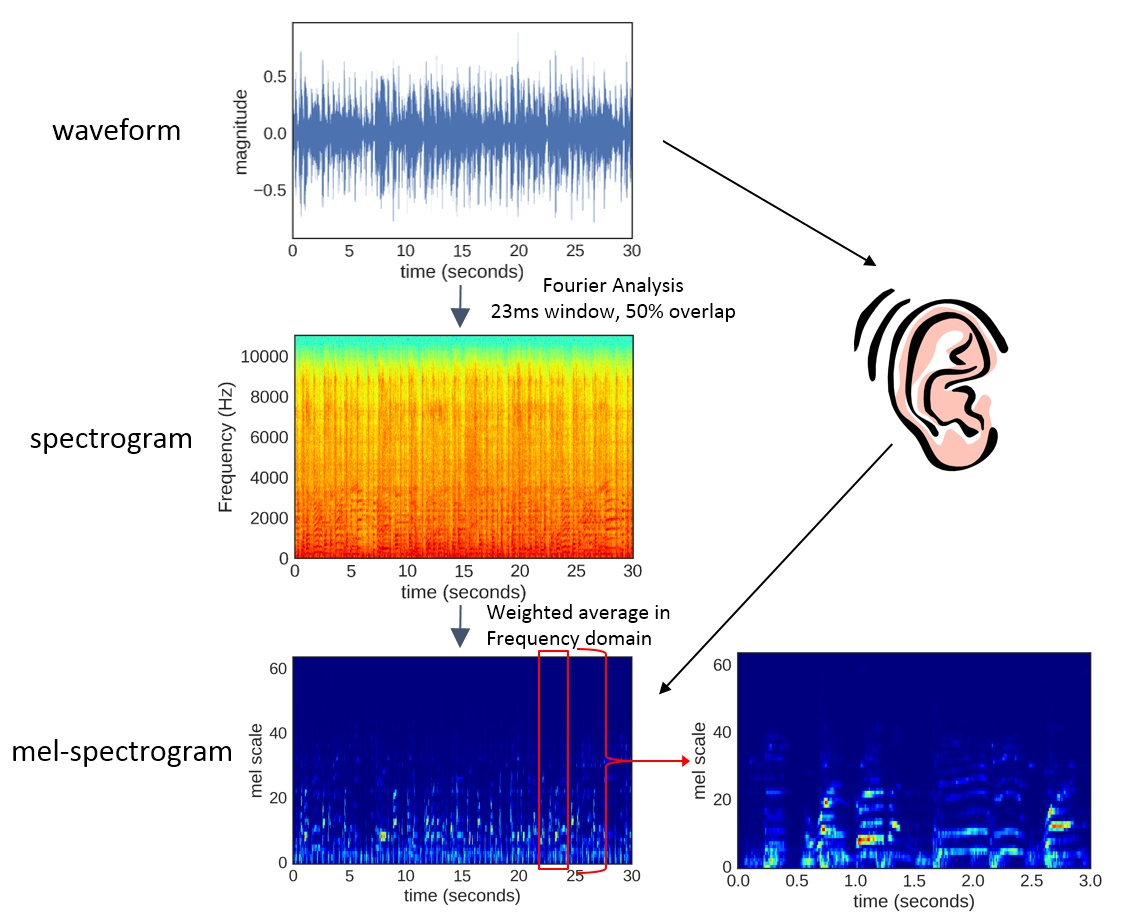}
	\caption{Convert waveform into mel-spectrogram and an example 3-second segment. Mel-spectrogram mimics how human ear works, with high precision in low frequency band and low precision in high frequency band. Note, the mel-spectrogram shown in the figures is already log transformed.}
	\label{fig: waveform_mel_spectrogram}
\end{figure}

Each music signal is first converted from waveform into mel-spectrogram $z_i$ using Librosa library with 23ms time window and 50\% overlap (figure \ref{fig: waveform_mel_spectrogram}). Then, the mel-spectrogram is log transformed to bring values at different mel-scale to the same range ($f(z_i) = ln(z_i + 1)$). Because mel-spectrogram is a biological-inspired representation \cite{picone1993signal}, it has a simpler interpretation than the PCA-whitening method used in \cite{lee2009unsupervised}.

\subsection{Network Architecture}
\begin{enumerate}
	\itemsep0em
	\item Input layer: 64 * 256 neurons, corresponds to 64 mel scales and 256 time windows(23ms, 50\% overlap).
	\item Convolution layer: 64 different 3 * 3 filters with a stride of 1. 
	\item Max pooling layer: 2 * 4.
	\item Convolution layer: 64 different 3 * 5 filters with a stride of 1.
	\item Max pooling layer: 2 * 4.
	\item Fully connected layer: 32 neurons that are fully connected to the neurons in the previous layer.
	\item Output layer: 10 neurons that are fully connected to neurons in the previous layer.
\end{enumerate}
For 2D layers/filters, the first dimension corresponds to the mel-scale and the second dimension corresponds to the time. All hidden layers use RELU activation functions, the output layer use softmax function, and the loss is calculated using cross-entropy function. Dropout and L2 regularization were used to prevent extreme weights. The model is implemented using Keras (2.0.1) with tensorflow as backend and trained on a single GTX-1070 using stochastic gradient descent.

\subsection{Training \& Prediction}
1000 music tracks (converted into mel-spectrogram) are evenly split into training, validation, and testing set with a ratio of 5 : 2 : 3. The training procedure is as following:
\begin{enumerate}
	\itemsep0em
	\item Select a subset of tracks from the training set. 
	\item Randomly sample a starting point and take the 3-second continuous segments from all selected tracks. 
	\item Calculate the gradients using back-propagation algorithm using the segments as input and the labels of the original music as target genres.
	\item Update the weights using the gradients.
	\item Repeat the procedure until classification accuracy on the cross-validation data set doesn't improve anymore. 
\end{enumerate}

During testing, all music (mel-spectrogram) are split into consecutive 3-second segments with 50\% overlap. Then, for each segment, the trained neural network predicts the probabilities of each genre. The predicted genre for each music is the genre with highest averaged probability.

\subsection{Calculate the filters learned by the CNN}
After training, all musics are split into 3-second segments with 10\% overlap. All the segments are then fed into the trained CNN and intermediate outputs are calculated and stored. Then, we estimated the learned filters using the following method:
\begin{enumerate}
	\itemsep0em
	\item Identify the range of input neurons (specific section of the input mel-spectrogram) that could activate the target neuron at a specific layer. E.g., $c_{i, j}^{(l)}$ indicates the neuron at location $(i, j)$ from the $l^{th}$ layer.
	\item Perform Lasso regression with the specific section of the mel-spectrogram (reshaped as a vector) as the regressors and the corresponding activations of the neuron $c_{i, j}^{(l)}$ as the target values.
	\item The fitted Lasso coefficients were reshaped to estimate the learned filters.
\end{enumerate}

\section{Results}

\begin{figure}
	\centering
	\includegraphics[width=0.7\linewidth]{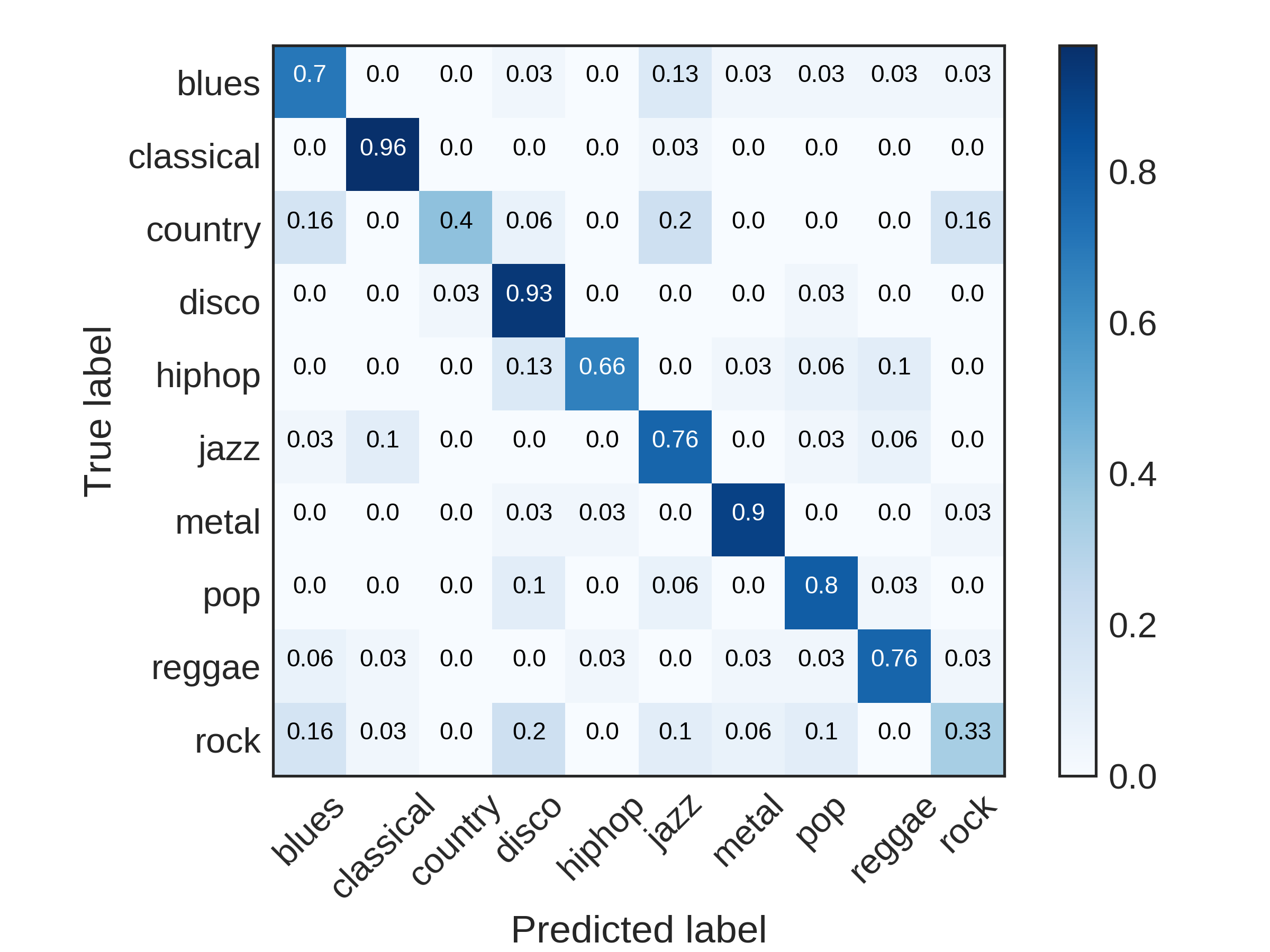}
	\caption{Confusion matrix of the CNN classification on testing set.}
	\label{fig: cnn_classification_results}
\end{figure}

To the best of our knowledge, the current model is the first to achieve human-level (70\%) accuracy in the 10-genre classification task (figure \ref{fig: cnn_classification_results}). It's 10\% higher than that obtained in \cite{tzanetakis2002musical} and classifies 5 more different genres than \cite{lee2009unsupervised} with similar accuracy. %Lee and colleagues achieves comparable accuracy using convolutional deep belief network but their model is only tested on a easier 5-genre classification task \cite{lee2009unsupervised}.

\subsection{Classification accuracies varies by different genres.}
From the confusion matrix (figure \ref{fig: cnn_classification_results}), we could see that the classification accuracy varies a lot across different genres. Especially, the accuracies for country and rock genre are not only lower than the current average but also lower than those from \cite{tzanetakis2002musical} (which has overall lower accuracy that our CNN). Because some important human-engineered features used in \cite{tzanetakis2002musical} are the long-term feature like beat and rhythm, this suggests country and rock music may have characteristic features (e.g., beat) that require longer time ($> 3$ seconds) to capture and 3s segments used in our CNN are not long enough. One future direction is to explore how to use CNN to extract long-term features for classification and one possibility is to use another down-sampled mel-spectrogram of the whole audio as input. Another explanation is that country and rock share more features with the other music genres and are more difficult to classify in nature. Nonetheless, expert advice probably is required to improve the classification accuracy on the country and rock genre. 

\subsection{CNN learns filters like spectro-temporal receptive field.}
%One way to characterize how human auditory system works is to calculate the spectro-temporal receptive field (STRF) of the neurons. Different STRF captures different stimulus patterns in the spectro-temporal space and the patterns captured by these STRF may be the features used to perform music genre classification and other tasks. 

\begin{figure}
	\centering
	\includegraphics[width=0.9\linewidth]{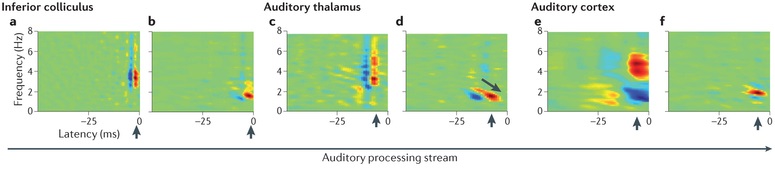}
	\caption{STRF obtained from physiological experiments. From left to right are the STRFs obtained from lower to higher auditory structures. Adapted from \cite{theunissen2014neural} with permission.}
	\label{fig: real_strf}
\end{figure}

\begin{figure}
	\centering
	\includegraphics[width=0.95\linewidth]{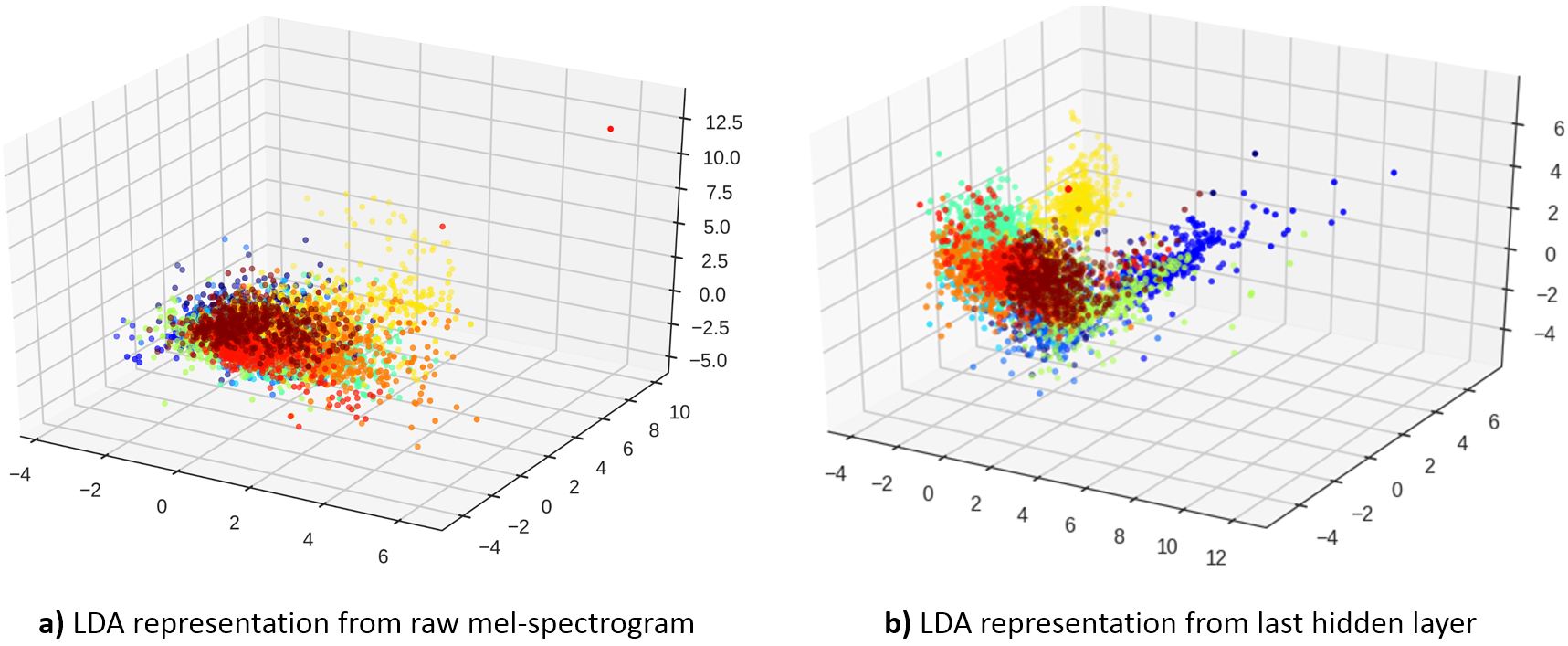}
	\caption{Comparison between the separability of the raw representation and last layer representation of the CNN of the testing data. The axes are the first three components when data is projected onto the directions obtained from linear discriminant analysis (LDA). using training data.}
	\label{fig: raw_last_layer_representation}
\end{figure}

Figure \ref{fig: cnn_classification_results} shows some filters learned by the CNN's 2nd max pooling layer and they're qualitatively similar to the STRF obtained from physiological experiments (figure \ref{fig: real_strf}). To visualize how these filters help classify the audios, we feed all the 3s segments from the testing set into the CNN and calculated the activations of the last hidden layer. After this non-linear transformation, most testing data points become linearly separable (figure \ref{fig: raw_last_layer_representation}). In contrast, the testing data points are much less separable when raw mel-spectrogram is used.

These results together show that the CNN learns filters similar to the spectro-temporal receptive field observed in the brain. These filters transform the original mel-spectrogram into a representation where the data is linearly separable. 

\section{Discussion}
By combining the knowledge from human psychophysics study and neurophysiology, we used the CNN in a "divide and conquer" way and classified the audio waveforms into different genres with human-level accuracy. The same technique may be used to solve problems that share similar characteristics, for example, music tagging and artist identification using raw audio waveform. With the current model, the genre of the music can be extracted efficiently with human-level accuracy and used as features for recommending music to the users.

\begin{figure}
	\centering
	\includegraphics[width=0.9\linewidth]{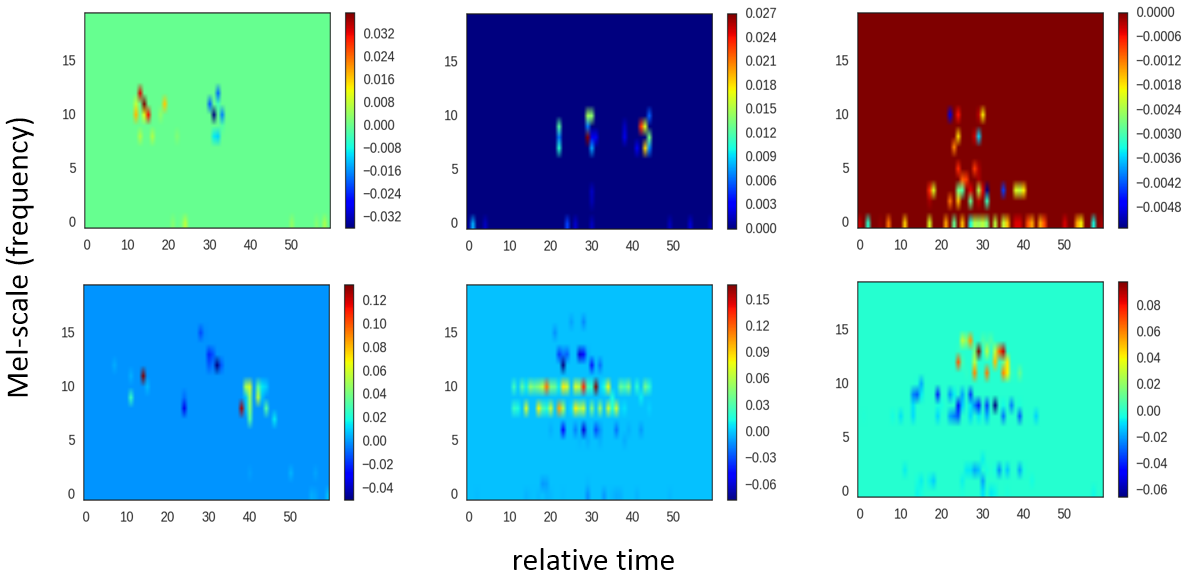}
	\caption{Filters learned by the CNN are similar to the STRF from physiological experiments. Mel scale corresponds to frequency and relative time corresponds to latency in figure \ref{fig: real_strf}. }
	\label{fig: filters_learned_cnn}
\end{figure}

\newpage
\bibliographystyle{unsrt}
\bibliography{cnn_mgc}

\end{document}